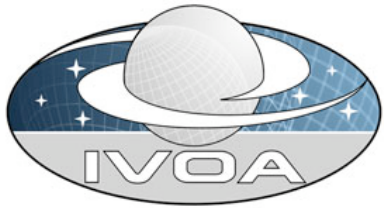

**I**nternational

**V**irtual

**O**bservatory

**A**lliance

# IVOA Identifiers
# Version 1.12

## IVOA Recommendation 2 March 2007

**This version:**
  http://www.ivoa.net/Documents/REC/Identifiers/Identifiers-20070302.html
**Latest version:**
  http://www.ivoa.net/Documents/latest/IDs.html
**Previous versions:**
  http://www.ivoa.net/Documents/PR/Identifiers/Identifiers-20060822.html
  http://www.ivoa.net/Documents/PR/Identifiers/Identifiers-20050302.html
  http://www.ivoa.net/Documents/PR/Identifiers/Identifiers-20040621.html
  http://www.ivoa.net/Documents/WD/Identifiers/Identifiers-20040209.html
  http://www.ivoa.net/Documents/PR/Identifiers/Identifiers-20031031.html
  http://www.ivoa.net/Documents/WD/Identifiers/Identifiers-20030930.html
  http://www.ivoa.net/Documents/WD/Identifiers/Identifiers-20030830.html
**Resource Registry Working Group:**
  http://www.ivoa.net/twiki/bin/view/IVOA/IvoaResReg
**Author(s):**
  Raymond Plante, Editor
  Tony Linde
  Roy Williams
  Keith Noddle

the IVOA Registry Working Group.

## Abstract

An IVOA Identifier is a globally unique name for a resource. This name can be used to retrieve a unique description of the resource from an IVOA-compliant registry. This document describes the syntax for IVOA identifiers as well as how they are created. An IVOA identifier has two separable components that can appear in two equivalent formats: an XML-tagged form and a URI-compliant form. The syntax has been defined to encourage global-uniqueness naturally and to maximize the freedom of resource providers to control the character content of an identifier.

## Status of this Document

This is a Proposed Recommendation. The first release of this document was 2004 February 9. It deprecates Proposed Recommendation: IVOA Identifiers v1.0.

Comments on this document are due 31 July 2004 for consideration in the next version of this document. They should be sent to registry@ivoa.net, a mailing list with a public archive. General discussion of related technology is also welcome on the Resource Registry wiki site.

This is an IVOA Proposed Recommendation made available for public review. It is appropriate to reference this document only as a recommended standard that is under review and which may be changed before it is accepted as a full recommendation.

A list of current IVOA Recommendations and other technical documents can be found at http://www.ivoa.net/Documents/.

## Acknowledgments

This document builds on the concept of a Uniform Resource Identifier as described in the IETF RFC 2396 by Berners-Lee, Fielding, Irvine, & Masinter [RFC 2396] and the subsequent IETF Internet Draft revisions [Berners-Lee et al. 2003].

This document has been developed with support from the National Science Foundation's Information Technology Research Program under Cooperative Agreement AST0122449 with The Johns Hopkins University, from the UK Particle Physics and Astronomy Research Council (PPARC), and from the European Commission's Sixth Framework Program via the Optical Infrared Coordination Network (OPTICON).

## Conformance-related definitions

The words "MUST", "SHALL", "SHOULD", "MAY", "RECOMMENDED", and "OPTIONAL" (in upper or lower case) used in this document are to be interpreted as described in IETF standard, RFC 2119 [RFC 2119].

The **Virtual Observatory (VO)** is general term for a collection of federated resources that can be used to conduct astronomical research, education, and outreach. The **International Virtual Observatory Alliance (IVOA)** is a global collaboration of separately funded projects to develop standards and infrastructure that enable VO applications.

## Syntax Notation

This document uses the Augmented Backus-Naur Form (ABNF) notation ([RFC 2234](https://www.ietf.org/rfc/rfc2234.txt)) to formally define syntax for identifier components. It references the following core ABNF syntax productions (defined in section 6.1): ALPHA, DIGIT.

## Contents

---

## 1. Introduction

There will be many occasions in Virtual Observatory applications where metadata will need to refer to some other data or concept unambiguously which is described elsewhere. In these cases, it is valuable to use a global identifier (ID) as a reference to the external item. An unambiguous reference within the entire IVOA community requires that the identifier be globally unique. Ensuring this uniqueness inevitably will require oversight by a moderating authority; however, a flexible framework can minimize the opportunity for duplicated IDs.

Many data providers in the VO have been creating and using IDs for a long time. Their choices of identifiers were made presumably to best fit the needs of the data. If an IVOA ID framework is to minimize the cost of adoption, then it needs to maximize the control of providers have to reuse the IDs they already have in place as well as create new IDs that are consistent with their overall organization. In addition, providers will need full flexibility over what an ID refers to. It could be a dataset that can be transmitted over the network, or a scientific instrument which cannot, or an abstract concept or organization.

Identifiers are very important to registries as they aid users in discovering data and services. In general, a registry stores descriptions of data and services in a searchable form, and it distinguishes them by a unique ID. Furthermore, when users encounter an ID, they should be able to go to a registry and find out something about the thing it refers to.

It is important to distinguish between two forms of referencing. The first is a reference to a specific instance of something. Typically, we think of this thing having a specific location; however, in the framework proposed here, we think of it as being held and managed by a specific organization even though its physical location may be undefined. The framework recognizes that entities like datasets do not always remain in the control of a single organization forever; thus, necessitating a second form of referencing that is location-independent--or more precisely, organization-independent.

When several copies of a dataset exist at several locations around the VO, one can refer to all of them collectively, deferring the choice of a particular instance until it is actually needed. Also, the curation of a dataset may be transferred from one organization to another; an organization-independent reference thus serves as a *persistent* pointer to data that can be resolved to a new location when it moves. This is very important to journal publishers that wish to refer to data in publications (whose useful life might be measured in decades) without worry that the references will become obsolete.

This proposal defines identifiers of the first type described above--that is, *organization-dependent identifiers*. Persistent, organization- and location-independent identifiers are *not* currently defined as part of this proposal, because it is not clear under what conditions it can be claimed that a resource is unchanged when its management is changed or duplicated from one organization to another. It is expected that a standard for organization-independent identifiers will be based on this proposal and the standard registry framework.

Nevertheless, these two forms of referencing are precisely what are addressed by the IETF standards for URIs [RFC 2396] and URNs [RFC 2141]. Thus, the framework proposed in this document builds directly on these standards. They provide an explicit mechanism for communities to add additional syntactic restrictions onto URIs and URNs and to define rules for interpreting them, all while still remaining compatible with the generic standards. This proposal uses this mechanism to enable the retrieval of descriptions of referenced items given their specific identifier and the resolving of organization-independent identifiers to specific ones.

## 1.1. Definitions

A **Uniform Resource Identifier (URI)** is defined by RFC 2396 as "a compact string of characters for identifying an abstract or physical resource" which complies with the syntax specification of that document (Berners-Lee et al. 1998). It can point to an actual retrievable resource (i.e. a URL is a type of URI); however, it need not. The use of the term URI in this document typically refers to an abstract identifier that *cannot* be used as a URL to retrieve things directly via HTTP. A **Uniform Resource Name (URN)** is a type of URI that serves as a "persistent, location-independent" identifier (RFC 2141).

In the Internet world, a **resource** is essentially anything that can be referenced by a URI. This document refines this definition by adding that it is describable by the generic resource metadata defined in the IVOA Recommendation on Resource Metadata (Hanisch et al. 2004, here on referred to as RM).

We also refer to organizations and providers in the sense that they are defined in the RM:

> An **organization** is a specific type of resource that brings people together to pursue participation in VO applications. Organizations can be hierarchical and range greatly in size and scope. At a high-level, it could be a university, observatory, or government agency. At a finer level, it could be a specific scientific project, space mission, or individual researcher. A **provider** is an organization that makes data and/or services available to users over the network.

Definitions of other types of resources, including data collection and service, are also defined in the RM, and are assumed by this document.

As discussed above, identifiers are critical to registries. A **registry** is a resource that stores metadata about other resources, including organizations, data collections, and services, and makes that information accessible through a set of services. Typically, one gets at this information either by doing a search against the metadata or through a look-up operation given an identifier. **Registration** is an operation that a provider carries out to tell a registry that a resource exists and can be referred to by a particular identifier; this is typically done through a registry service. An IVOA-compliant registry is a registry that implements the minimum, IVOA standard registry services (currently under development), including description look-up by ID. Upon registering a resource, its provider becomes known as the resource's **publisher**.

## 1.2. Selected Requirements

This proposal follows on various requirement studies for VO identifiers and registries in general (e.g. NVO ID requirements). This section highlights a few of the important ones that guided the design of the ID framework.

1.  A single framework should be used to identify anything a VO application can refer to, including organizations, projects (mission/telescope), data collections, and services.

2.  It should be easy to compare two instances of an identifier to determine if they refer to the same object.

3.  It should be possible to use an identifier to access a unique description of the resource it identifies.

4.  The framework should maximize the freedom of data providers to choose identifiers for resources and collections under their control.

## 2. Overview

The IVOA identifier framework supports two forms of an identifier: an XML form and a URI form. An example of the XML form looks like this:

```
<ResourceID>
   <AuthorityID>adil.ncsa</AuthorityID>
```

```
        <ResourceKey>surveys/96.JC.01</ResourceKey>
    </ResourceID>
```

The same identifier can be expressed as a URI:

```
ivo://adil.ncsa/surveys/96.JC.01
```

In both forms, the identifier has two components: an authority ID and a resource key. The former establishes a namespace within which the rest of the ID, the resource key, can be considered unique.

Identifiers are considered case-insensitive; however, the preferred rendering of character case in the ID is determined when its resource is registered.

## 3. Specification

An **IVOA Resource Identifier** (or *IVOA identifier* or *IVOA ID* for short) is a globally unique reference to a resource represented in a compact, ASCII-text format, as prescribed in this document. This document defines two formats: an XML format and a URI format. An IVOA identifier MUST always refer to a resource that has been registered with an IVOA-compliant registry; that is, it should be possible to use the ID to get a description of the resource from a compliant registry somewhere in the VO environment.

### 3.1. Identifier Components and their Syntax

An IVOA ID has two required logical components: an *authority identifier* and a *resource key*.

**3.1.1. The Authority Identifier**

A **naming authority** is an organization (usually a provider) that has been granted the right by the IVOA to create IVOA-compliant identifiers for resources it registers. (See Creating Identifiers for details on how this right is granted.) The naming authority creates IDs within the scope of one or more authority identifiers.

An **authority identifier** is a compact string of ASCII text that defines a globally unique namespace controlled by a single naming authority. The authority ID string is a compliant URI authority component (RFC 2396, section 3.2) with the following restrictions:

- it must be at least three (3) characters long,
- it must begin with an alpha-numeric character,
- it must not contain URI-escaped sequences, and
- it must not contain any of these reserved characters, "!", ";", ":", "@", "&", "$", nor ","

> **Note:**
> While the syntax for the authority ID allows it to look just like a DNS hostname, current convention discourages this practice to avoid the suggestion that an IVOA Identifier can be resolved like a URL. As of this writing, the convention of the National Virtual Observatory (NVO) is hierarchical naming that combines the publishing organization name with the project or archive (e.g. "adil.ncsa") while leaving out fields like ".edu" or ".org". In the AstroGrid project, the convention is to use a DNS name in reverse order (e.g. "org.astrogrid.www"); this practice has the advantage of reducing the probability that two organizations will want to use the same authority ID.

> **Tip:**
> The URI syntax rules for an authority (RFC 2396, section 3.2) already disallow the following characters:
> - control characters,
> - the space character,
> - delimiters, "<", ">", "#", "%",
> - double quotes and forword single quotes ("`"),
> - URI reserved characters, "/", "?",
> - selected non-alphanumeric characters, "{", "}", "|", "\", "^", "[", "]".
>
> Consquently, they cannot appear in an authority ID either.
>
> To make an authority ID more recognizable, the following characters are discouraged from being a part of the authority ID: "~", "*", "'", "(", and ")". It is also recommended that authority IDs avoid the multiple, sequential occurrences of periods, ".".

In ABNF notation, the syntax for an authority ID is as follows:

```
authorityid = alphanum 2*unreserved
```

```
alphanum    = ALPHA / DIGIT

reserved    = "?" / ";" / ":" / "@" /
              "!" / "&" / "$" / ","

unreserved  = alphanum / mark / discouraged

mark        = "-" / "_" / "."

discouraged = "~" / "*" / "'" / "(" / ")"
```

A naming authority is allowed to control multiple authority IDs to organize related resources into different namespaces. For example, an organization may choose to control two authority IDs, one for research-related resources and one for education/outreach resources, even though they are all maintained by the same organization and perhaps made available through the same machine.

VO applications should be case-insensitive when handling authority IDs (see "Comparing Identifiers" below). In practice, applications are encouraged to present identifiers using all lower-case characters.

**3.1.2. Resource Key**

A **resource key** is a localized name for a resource that is unique within the namespace of an authority ID. The naming authority creates keys for its namespaces and has complete control of their forms beyond the syntax constraints specified here.

A resource key must conform to the syntax of a URI path component (RFC 2396, section 3.2); that is, it is a slash ("/") delimited ASCII string. In addition, it must not contain any of the other reserved characters listed in section 3.1.1.

In ABNF notation, the syntax for a resource key is as follows:

```
resourcekey = segment *( "/" segment)
segment     = *unreserved
```

Naming authorities are discouraged from creating segments matching either "." or "..". Empty segments, resulting in two or more consecutive slashes or a trailing slash, is also discouraged. As described in the section "Comparing Identifiers", such segments do not have the special meaning they have in traditional file system pathnames; that is, a resource key cannot be transformed to remove any "." or ".." segments and still reference the same resource. Such segments are interpreted as a literal component of the path.

> **Note:**
> The reserved characters may be employed in the future for adapting IVOA identifiers into various applications. This might include differentiating additional components of an identifier beyond the authority ID and resource key or otherwise delimiting an ID within some other string.

The naming authority is free to create a resource key that suggests something about the resource it refers to. Any meaning that is suggested by the resource key is intended only for human consumption. The character content of a resource key is not semantically machine-interpretable within the context of the IVOA as defined by this document. Any information about the resource, apart from whether it is same exact resource (i.e. instance or copy) as the resource referred to by another identifier (see "Comparing Identifers"), can only be determined by examining the resource metadata.

The presence of a resource key is optional. An identifier that contains only an authority ID refers to the registered namespace itself and, indirectly, the organization that acts as the naming authority for that namespace.

VO applications should be case-insensitive when handling a resource key (see "Comparing Identifiers" below). In practice, the preferred use of case is set by the rendering of the key by the naming authority when the resource is registered. This may contain one or more capital letters to improve reading.

### 3.2 Identifier Formats

The identifier components can be combined into two equivalent formats: an XML-tagged form and a URI-compliant form.

**3.2.1. XML Format**

An IVOA identifier rendered in XML can be described as an XML complex type as defined by the XML Schema specification (XMLSchema-0). This type contains two child elements. The first element, which must be present and is non-repeatable, is named **AuthorityID**, and its content is a string that conforms to the syntax of an authority ID. The second, non-repeatable element is named **ResourceKey**, and its content is a string that conforms to the syntax of a resource key.

Appendix A lists an XML Schema definition which includes the XML element, Identifier, that is defined to be an IVOA identifier for a registered resource. Other schemas may import and use this element directly wherever this general metadata concept is useful. If a more precise meaning is needed, e.g. if "PublisherID" is meant to refer to a resource's publisher, then a schema can create a new element whose type is "IVOAidentifier".

**3.2.2. URI Format**

There will be occasions when an application needs to encode an identifier in a format where the XML encoding defined in the previous section cannot be readily used but where a simple string can. This would include a non-XML format, such as in a FITS keyword. A number of XML-based metadata handling systems also treat identifiers as strings (e.g. Resource Description Language, Dublin Core, Open Archives Initiative). In some cases, identifiers are encoded as XML attributes. To enable greater compatibility with other metadata technologies, this document also defines a URI format for an identifier.

This specification defines a new URI scheme called "ivo." A URI that uses this scheme signals that:

- the identifier and the resource it refers to have been registered with an IVOA-compliant registry.
- the URI can be parsed into an authority ID and a resource key according to the syntax defined in this section.

The URI form of an identifier starts with the "ivo" scheme string and is followed by a colon and two slashes. The double slashes reflect the URI convention that a URI refers to a remote resource. After the double slashes is the authority id which must match the syntax given in section 3.1.1. After that is a slash followed by the resource key, which must match the syntax given in section 3.1.2. Thus, the authority ID is separated from the resource key by the first slash after the double slashes.

In addition to the above syntax rules, the URI format reserves the question mark ("?") and the pound sign ("#") as "stop" characters: that is, when one of these two characters appears in a string beginning the "ivo" scheme, all characters before the first question mark or pound constitute the complete IVOA resource identifier, and all characters after and including the "stop" character are to be ignored when handling the string as an IVOA resource identifier.

> **Note:**
> The purpose of a stop character is to allow an IVOA resource identifier to be concatenated with another string, creating a type of augmented or annotated identifier for some application-specific purpose. This specification does not restrict in any way what comes after the stop character or what the augmented identifier might be used for. A common use, however, is to support globally unique *dataset identifiers*. Typically, the portion before the stop character is the IVOA identifier of registered data collection. The remaining portion, then, refers to a specific dataset within that collection which is *not* registered. A future version of this specification may adopt to this common practice by standardizing the syntax for identifiers that refer to entities which are not registered in an IVOA registry.

To simplify comparisons of identifiers in URI format (see "Comparing Identifiers"), case-insensitive variations on "ivo" as the scheme shall be considered equivalent; however, use of variations other than the all lower-case form are strongly discouraged.

In ABNF format, the URI form is defined as:

```
ivo-scheme = ("i" / "I") ("v" / "V") ("o" / "O")
uri-form   = ivo-scheme "://" authorityid ?( "/" resourcekey )
```

It is important to note that an IVOA resource identifier in URI format is not a URL. The ivo scheme does not imply a transport protocol by which the resource may be accessed. Agents, in general, should not depend on implicit mappings between IVOA resource identifiers and URLs (e.g. replacing the ivo scheme with the http scheme) to derive a URL for the resource. Resource publishers, however, may support such a mapping between identifiers and URLs that they manage; in this case; agents should only assume the mapping applies within the domain of the publisher.

## 3.3 Creating Identifiers

An important aim of the process for creating identifiers is to ensure uniqueness. In the context of IVOA resource identifiers, "unique" means that a given identifier MUST NOT refer to two different resources at any instant. Furthermore, the identifier SHOULD refer to one logical resource over all time; that is, IVOA identifier should not be reused. The description of the resource referred to by the identifier, however, MAY change over time. What constitutes the difference between re-use of an identifier and an update of its description is left up to the discretion of the publisher.

Another aim of the identifier creation process is to trace the delegation of authority over the identifier. In principle, the right to create identifiers is granted to naming authorities by the IVOA via a service defined among the standard registry services. In practice, an identifier is created by a person or organization when registering a resource. Thus, only recognized naming authorities (or persons representing such organizations) may register new resources.

The exact details of the service used to become a recognized naming authority is not described here; however, such a service has the following requirements:

- The service approves an organization as naming authority by associating one or more authority identifiers with the organization.

- The organization desiring recognition as a naming authority must be registered with a registry; that is, a registry must contain a standard IVOA description for that organization. This registered description must be accessible to the service.

    It is possible that an organization may become registered as part of the process of requesting an authority ID. Equivalently, registering an organization may result in an implicit request for a particular authority ID.

- If the service approves the request, the authority ID is considered controlled by the organization.

- In processing the request for an authority ID, the service must check to ensure that the ID is not already controlled by another organization. This is done by doing a by-identifier search of a global registry for the requested ID. If the authority identifier is already registered, then request is denied.

The registry service specification or a particular implementation may impose additional requirements for approving a request for an authority ID. Note, however, that it is intended that the requirements determining who may become a naming authority should not be overly restrictive, nor should it take much more time to obtain approval than what it takes to determine if the authority ID is already taken.

Once an organization is recognized as a naming authority, it is free to register any number of resources with identifiers having an authority ID that they control. No other organizations may create identifiers with an authority ID it does not control. The naming authority has full control over the creation of a resource key as long as it conforms to the syntax and uniqueness constraints described in this specification.

It is the responsibility of the registry that accepts new resource descriptions to ensure that new descriptions are not associated with identifiers already referring to other resources. The mechanisms used to ensure this are not described here.

Since IVOA resource identifiers refer to a resource that has been registered (see definition), providers and software agents should avoid referring to unregistered resources via an IVOA identifier. In practice, there may be some time lag between when a resource is available and when it is actually published; in this case, any use of the identifier should based on the expectation that the resource will be registered soon.

### 3.4 Comparing Identifiers

An important use of identifiers is comparing two instances to determine if they refer to the same resource. This will most commonly occur when using an identifier to look up the associated resource description in a registry.

Two resource identifiers are guaranteed to refer to the same resource if they have identical authority IDs and identical resource keys. A pair of components is considered identical if a *case-insensitive*, character-by-character comparison indicates they are identical. Apart from a transformation to handle case-insensitive comparisons, no other normalizing transformations shall be necessary to test if two resources are IDs refer to the same resource.

> **Note:**
> The case-insensitive string comparison test on two identifiers in URI format is equivalent to parsing to the identifiers into their two components and comparing them.

In general, the string-based comparison of identifiers described cannot determine definitively if two identifiers refer to different resources. While it is not intended that a single resource be registered multiple times with different identifiers, it is not disallowed by this specification. In particular, it is possible that two resources with different identifiers may be mirrors of each other; such a relationship can only be determined by examining the metadata contained in the descriptions associated with each identifier.

### Appendix A: An XML Schema for Identifiers

```
<?xml version="1.0" encoding="UTF-8"?>
<xs:schema xmlns:xs="http://www.w3.org/2001/XMLSchema"
           xmlns:vid="http://www.ivoa.net/xml/VOIdentifier/v1.1"
           targetNamespace="http://www.ivoa.net/xml/VOIdentifier/v1.1"
           elementFormDefault="qualified" attributeFormDefault="unqualified">

  <xs:annotation>
    <xs:documentation>Version 0.1</xs:documentation>
    <xs:documentation>
      This schema defines the XML format for IVOA Identifiers as
```

```
      specified in the IVOA Identifiers Working Draft, Version 0.1.
    </xs:documentation>
    <xs:documentation>
      History:
    </xs:documentation>
  </xs:annotation>

  <xs:element name="Identifier" type="vid:IVOAidentifier">
    <xs:annotation>
      <xs:documentation>
        a global, IVOA-compliant identifier that refers
        unambiguously to a resource.
      </xs:documentation>
    </xs:annotation>
  </xs:element>

  <xs:complexType name="IVOAidentifier">
    <xs:sequence>
      <xs:element ref="vid:AuthorityID" />
      <xs:element ref="vid:ResourceKey" minOccurs="0"/>
    </xs:sequence>
  </xs:complexType>

  <xs:element name="AuthorityID" type="vid:AuthorityIDType">
    <xs:annotation>
      <xs:documentation>the identifier a namespace under the control of
         a single naming authority</xs:documentation>
    </xs:annotation>
  </xs:element>

  <xs:element name="ResourceKey" type="vid:ResourceKeyType">
    <xs:annotation>
      <xs:documentation>the identifier a namespace under the control of
         a single naming authority</xs:documentation>
    </xs:annotation>
  </xs:element>

  <xs:simpleType name="AuthorityIDType">
    <xs:restriction base="xs:string">
      <xs:pattern value="[\w\d][\w\d\-_\.~\*'\(\)\+=]{2,}"/>
    </xs:restriction>
  </xs:simpleType>

  <xs:simpleType name="ResourceKeyType">
    <xs:restriction base="xs:string">
      <xs:pattern value="[\w\d\-_\.~\*'\(\)\+=]+(/[\w\d\-_\.~\*'\(\)\+=])*"/>
    </xs:restriction>
  </xs:simpleType>

  <xs:simpleType name="IVOAIdentifierURI">
    <xs:restriction base="xs:anyURI">
      <xs:pattern value="ivo://[\w\d][\w\d\-_\.~\*'\(\)]{2,}(/[\w\d\-_\.~\*'\(\)]+(/[\w\d\-_\.~\*'\(\)])*)"/>
    </xs:restriction>
  </xs:simpleType>

</xs:schema>
```

## Appendix B: Changes from Previous Versions

### B.1 Changes from v1.10

- Moved "!" from the discouraged list of characters to the reserved list, thereby disallowing its inclusion in IVOA identifiers.
- Clarified the list of characters disallowed in an authority ID (section 3.1.1) by:
    - explicitly disallowing URI-escaped sequences
    - listing as reserved characters only those characters that are allowed by the URI spec but disallowed by this one.
    - Listed in a tip box the characters that are disallowed by the URI spec.
  As before, the definition of the resource key (section 3.1.2) refers to the same list of reserved characters as those disallowed.
- Fixed numerous links and references

### B.2 Changes from v1.0

- The prohibition of using "+" and "=" within Identifier components has been dropped.
- Recommendations for authority ID strings have been updated to match current practice in AstroGrid and the NVO.
- In the example schema in App. A, the namespace was altered to conform with IVOA conventions. A correction was also made to the allowed pattern for AuthorityIDType to properly comply with the XML specification defined in section 3.2.1.

- various clarifications based on reviewer comments

### B.3 Changes from v0.1

- Resource key is now required except when referring to a naming authority itself (3.1, 3.1.2).
- support for DNS-like authority IDs clarified (3.1.1).
- added role of # and ? as "stop" characters in URI form (3.2.1).
- dropped non-binding Appendix B: Recommended Mechanism for becoming a Naming authority.